\documentstyle[12pt,aaspp4]{article}

\slugcomment{Accepted For Publication in ApJ Letters}

\begin{document}

\title {A wide-field {\em K}--band survey. The
luminosity function of galaxies.}

\author {J. P. Gardner$^{1,2}$, R. M.
Sharples$^{1}$, C. S. Frenk$^{1}$, \& B. E. Carrasco$^{3}$
\\$^1$University of Durham, Physics Dept., South Road, Durham DH1
3LE, ENGLAND \\ $^2$Laboratory for Astronomy and Solar Physics,
Code 681, Goddard Space Flight Center, Greenbelt MD 20771 \\$^3$INAOE,
Apdo Postal 216 y 51, Puebla, CP 72000, MEXICO \\ E-mail addresses:
gardner@harmony.gsfc.nasa.gov; r.m.sharples@durham.ac.uk;
c.s.frenk@durham.ac.uk; bec@tonali.inaoep.mx}

\begin{abstract}

We present the first determination of the near-infrared $K-$band
luminosity function of field galaxies from a wide field $K-$selected
redshift survey. The best fit Schechter function parameters are
$M^* = -23.12 +5log(h)$, $\alpha = -0.91$, and $\phi^* = 1.66 \times
10^{-2} h^3 ~ Mpc^{-3}$. We estimate that systematics are no more
than $0.1 mag$ in $M^*$ and $0.1$ in $\alpha$, which is comparable
to the statistical errors on this measurement.

\end{abstract}


\section{Introduction}

The luminosity function of galaxies is central to many problems in
cosmology, including the interpretation of faint number counts.
Because of this, the faint end slope of the luminosity function is
currently the subject of much debate. Most measurements of the
optical luminosity function of field galaxies show a flat slope,
corresponding to $\alpha \approx -1.0$ in the Schechter (1976)
parameterization (Efstathiou, Ellis \& Peterson 1988; Loveday et
al.\ 1992; Lin et al.\ 1996; but see Marzke et al.\ 1994). Deep
field galaxy surveys, on the other hand, detect a very steep slope
for the faint end of the optical number count relation, at the
point where the relation could be dominated by the faint end slope
of the local luminosity function (Tyson 1988; Lilly, Cowie \&
Gardner 1991; Metcalfe et al.\ 1996; but see also Cowie et al.\ 1996).

The near-infrared provides several advantages over the optical for
statistical studies of galaxies. The K--corrections due to the
redshift of the spectral energy distribution are smooth, well
understood, and nearly independent of Hubble type; the expected
luminosity evolution is also smooth. The $K-$band is dominated by
near-solar mass stars which make up the bulk of the galaxy. The
absolute $K$ magnitude is a measure of the visible mass in a galaxy,
and thus the $K-$band luminosity function is an observational
counterpart of the mass function of galaxies. Previous determinations
of the local $K-$band luminosity function have suffered from small
sample size or color-dependent incompleteness (Mobasher, Sharples
\& Ellis 1993; Glazebrook et al.\ 1995), while surveys conducted
at fainter levels are more appropriate for studying the evolution
of galaxies (Songaila et al.\ 1994; Cowie et al.\ 1996). In general,
studies of galaxy evolution through number counts, colors, redshift
distributions and clustering properties, all require an understanding
of the local population of galaxies for interpretation of the faint
end data.

We have conducted a photometric and spectroscopic survey of galaxies,
observed in the near-infrared and optical with linear detectors,
and have obtained spectroscopic redshifts for a sample of galaxies
selected in the near-infrared. We present here the $K-$band luminosity
function. Results of the photometry were presented in Gardner et
al.\ (1996; Paper I), and Baugh et al.\ (1996; Paper II). Future
papers in this series will present the catalog, an analysis of the
redshift and color distributions of the galaxies, the bivariate
optical-near infrared luminosity function of galaxies, and an
analysis of the star counts.

\section{The Luminosity Function}

The luminosity function of galaxies is the volume density of galaxies
as a function of their absolute magnitude. Field galaxy surveys
such as this one are typically magnitude limited, and the galaxy
distribution has structure along the line--of--sight of the survey.
Several methods have been developed to determine the luminosity
function from a redshift survey, avoiding systematics due to
clustering. Reviews are presented in Binggeli, Sandage \& Tammann
(1988) and Efstathiou et al.\ (1988; hereafter EEP).

The traditional method for determining the luminosity function from
a magnitude limited field galaxy redshift survey is to sum over
the inverse of the maximum volume within which each galaxy could
have been detected (Felton 1977), but this method is biased due to
clustering of the galaxies. Sandage, Tammann \& Yahil (1979;
hereafter STY) developed a maximum likelihood technique for fitting
a parametric form, in which the effects of clustering cancel out
on the assumption that the luminosity function does not depend on
position. EEP developed the step-wise maximum likelihood method
(hereafter SWML) in which the data are binned, but the results are
relatively insensitive to clustering and no parametric form is
assumed. Other workers extended these methods to include the effects
of photometric errors (Loveday et al.\ 1992), redshift errors
(SubbaRao et al.\ 1996), peculiar motions (Schechter 1976),
substantially incomplete data sets (Isobe \& Feigelson 1992) and
the use of likelihood to measure the goodness-of-fit of a parametric
form (Yahil et al.\ 1991).

\subsection{The Data}

\begin{table}

\caption{The dependence of the Schechter parameters on
cosmological geometry and evolutionary corrections.}

\begin{tabular}{ccccc}
$q_0$  & corrections & $M^* - 5log(h)$ & $\alpha$ & $\phi^* ~ h^{-3}$     \\
$0.5$  & K \& E      & $-23.12$        & $-0.91$  & $1.66 \times 10^{-2}$ \\
$0.5$  & K only      & $-23.37$        & $-1.03$  & $1.82 \times 10^{-2}$ \\
$0.02$ & K \& E      & $-23.30$        & $-1.00$  & $1.44 \times 10^{-2}$ \\
$0.02$ & K only      & $-23.51$        & $-1.09$  & $1.50 \times 10^{-2}$ \\
\end{tabular}

\label{geomtab}
\end{table}

We have conducted a spectroscopic redshift survey of galaxies
selected on the basis of their $K-$band flux in an area of
approximately 4.4 square degrees, from within a larger photometric
survey of 10 square degrees. The photometric observations in two
fields of roughly equal area were made in the $K$ band with a HgCdTe
NICMOS3 detector in 1994 June on the Kitt Peak National Observatory
$1.3 m$ telescope, and in the $B$, $V$, and $I$ bands with a $2048^2$
CCD camera in 1995 June on the KPNO $0.9 m$ telescope. In 1996 May,
we used the $4.2 m$ William Herschel Telescope on La Palma with
the Autofib-2 fiber positioner and the WYFFOS spectrograph, to
obtain spectra of 567 objects selected at $K<15$. Approximately
75\% of the spectroscopic observations were made in the NGP field,
and the remainder were made in the NEP field (see Paper II). Objects
within the photometric sample were selected for spectroscopy on
purely geometrical criteria determined by the characteristics of
Autofib-2. There is a small bias against interacting galaxies,
since fibers could not be placed closer than 30 arcsec apart. The
data reduction was done with the wyfred package written by Jim
Lewis at the Royal Greenwich Observatory, and the redshift
identification was done with software developed by Karl Glazebrook.
We obtained good identifications and redshifts of 465 galaxies in
the sample, and less certain redshifts for an additional 45 galaxies.
The latter have spectra with poor signal-to-noise, poor sky
subtraction, or only a single significant line or break. Three
objects were identified spectroscopically as stars. The remaining
54 galaxies were unidentified, giving a completeness of 90\%.

\subsection{Determination of the Luminosity Function}

\begin{figure}

\plotone{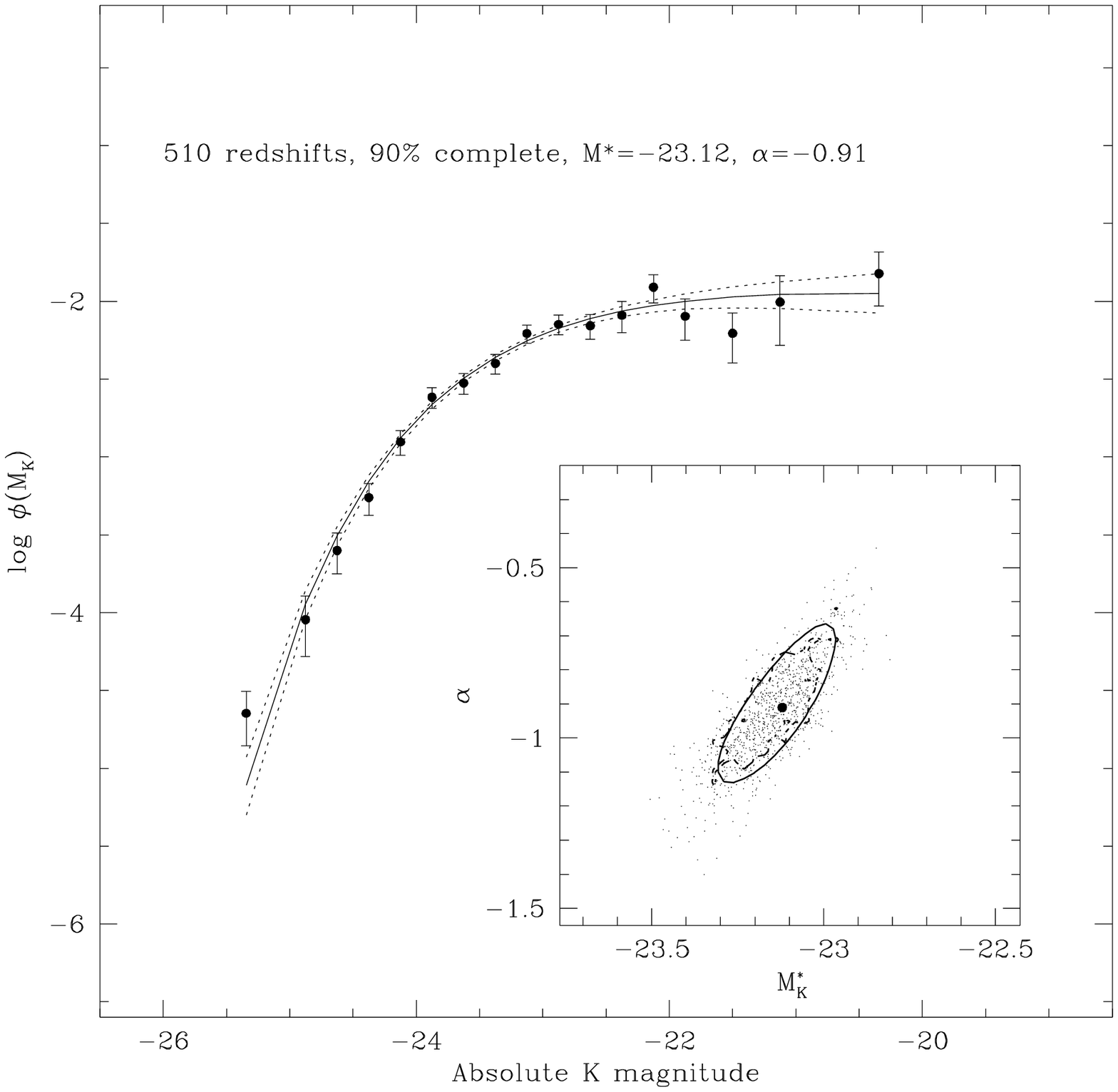}

\caption{The differential $K-$band luminosity function of galaxies.
The points and their errors were determined from our data using
the SWML method of EEP. The solid line is the best fit Schechter
function determined using the STY maximum likelihood method. The
dashed lines show the effect of varying the parameters of the fit
by $\pm 1\sigma$, as determined from the error ellipse. Inset are
the error ellipse on the Schechter parameter fit to the luminosity
function, and the results of 1000 Monte Carlo simulations of our
survey parameters. These simulations were binned as 0.03 in $M^*$
and 0.03 in $\alpha$, and a contour containing 68\% (i.e.\ $1\sigma$)
of the points in the binned data is shown as a dashed line.}

\label{klffig}

\end{figure}

We calculated the luminosity function from our data using the SWML
method. The results are plotted in Figure~\ref{klffig}. We determined
the variances using the constraint given in EEP, with $M_{fid} =
-23.5 +5log(h)$ and $\beta = 1.5$. We used the STY maximum likelihood
method to determine the best fit Schechter function parameters,
$M^* = -23.12 + 5 log(h)$, and $\alpha = -0.91$, and this function
is plotted as a solid line in Figure~\ref{klffig}, with the error
ellipse in $M^*$ and $\alpha$ plotted in the inset. The dashed
lines on either side of the Schechter function in Figure~\ref{klffig}
show the effect of varying $M^*$ and $\alpha$ by $\pm 1\sigma$. As
a check of our error determinations, we ran 1000 Monte Carlo
simulations of the survey. Using a number count model, we assigned
redshifts and absolute magnitudes to a mock sample of 510 galaxies
selected at $K<15$, and we determined the Schechter parameters
using the STY method. These 1000 simulations are plotted in the
inset to Figure~\ref{klffig}. The contour enclosing 68\% of the
points agrees approximately with the error ellipse determined from
the likelihood. The mean parameter values from the 1000 simulations
differed from the inputs by $-0.03 mag$ in $M^*$, and $-0.01$ in
$\alpha$, and this may be taken as an indication of systematic
errors in the techniques used, and of the inaccuracy in the number
count model.

Errors in the photometry affect the determination of the luminosity
function. Our $K-$band photometry is accurate to about $0.1 mag$
at the selection limit of $K=15.0$. The steep slope of the number
counts at this magnitude will result in more galaxies scattering
into the sample from fainter magnitudes, than scattering out of
the sample. We investigated this effect with 1000 Monte Carlo
simulations. We created a mock catalog limited at $K<15.5$, added
Gaussian noise with $1 \sigma = 0.10$, selected a new $K<15.0$
catalog, and determined the corresponding Schechter function
parameters. The mean results of these simulations differed from
the input parameters by only $-0.04 mag$ in $M^*$ and $-0.04$ in
$\alpha$.

K--corrections are relatively independent of Hubble type in the
$K-$band, but nonetheless we used the method of Eales (1993),
described by Gardner (1996), to determine the types. We adopted
models for the $B-V$, $V-I$ and $I-K$ colors as a function of
redshift from five Bruzual \& Charlot (in preparation; hereafter
GISSEL96) solar metallicity models with different star formation
histories. A least-squares fit to the colors of the galaxies provided
the rest-frame SED, which was convolved with the filter response
function to obtain the rest-frame absolute $K$ magnitude. Our goal
is to measure the zero redshift luminosity function of galaxies,
and so we included the effects of passive evolution (that is,
E--corrections), in our fits and in the determination of the
luminosity function. The best-fit Schechter parameters, including
K--corrections, but ignoring the effects of passive evolution are
given in Table~\ref{geomtab}. The difference in $M^*$ measured in
these two cases reflects the evolution of an elliptical galaxy from
the median redshift of the survey (z=0.14) to the present, which
is $\Delta M_K=-0.17$ in the GISSEL96 model.

Throughout this paper we have used a cosmology in which $H_0 = 100
h~km~s^{-1}~Mpc^{-1}$, $q_0=0.5$, and $\Lambda=0$. Varying $q_0$
affects the calculation of absolute magnitude and the evolutionary
model. The best-fit Schechter functions for $q_0 = 0.02$ are listed
in Table~\ref{geomtab}. The difference in $M^*$ between the flat
and open cosmologies may be compared with $\Delta M_K=-0.13$, which
is the difference in the absolute magnitudes of an elliptical galaxy
at the survey median redshift when calculated in the two different
cosmologies.

\subsection{The normalization $\phi^*$}

\begin{figure}

\plotone{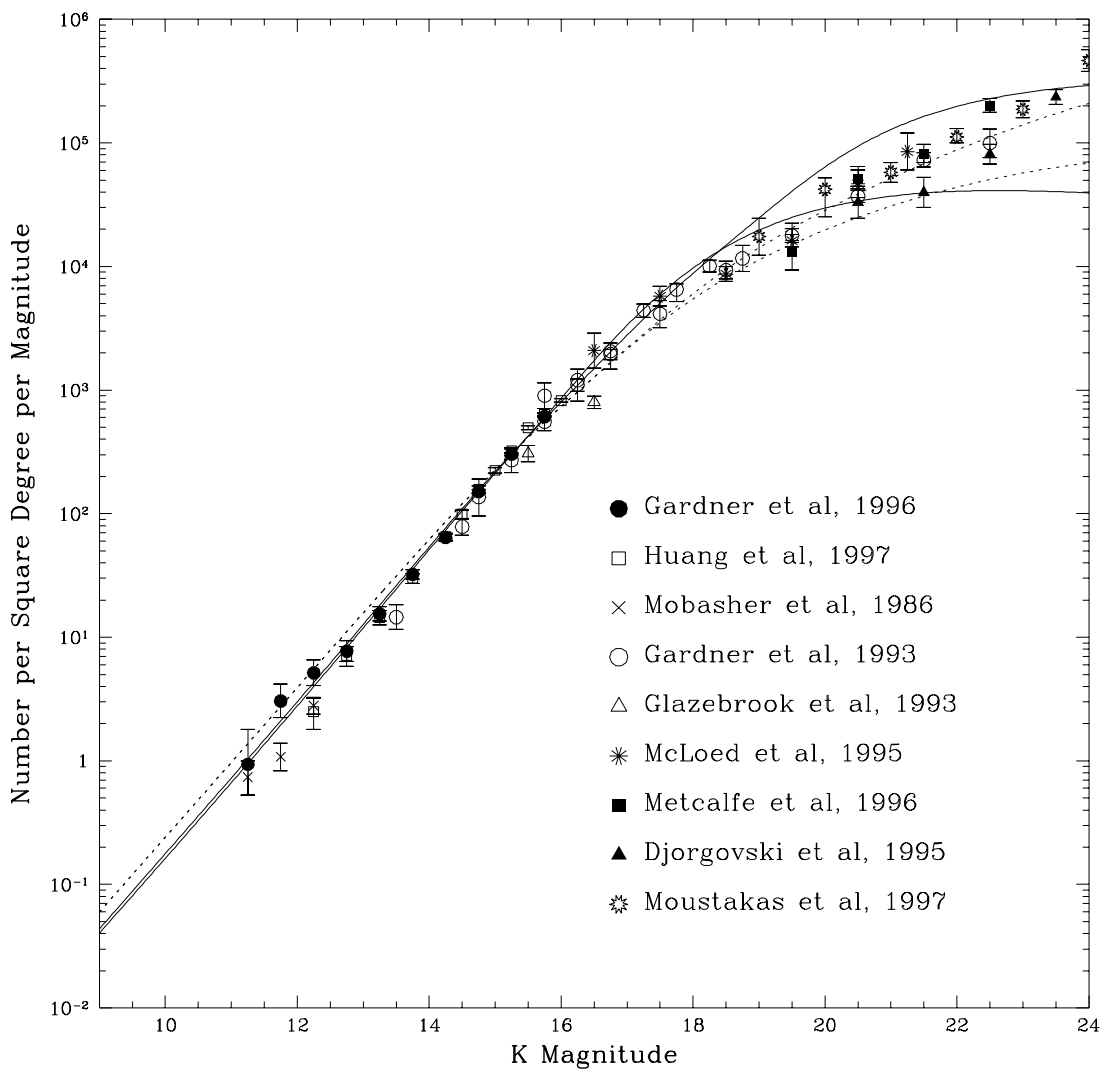}

\caption{The $K-$band number counts compared with models based upon
our estimated luminosity functions. The solid lines include the
effects of passive evolution, while the dotted lines include only
K--corrections. The higher line in each case is for $q_0 = 0.02$,
while the lower lines are for $q_0=0.5$.}

\label{fignc}

\end{figure}

The STY and SWML methods determine the shape of the luminosity
function, but not its normalization ($\phi^*$ in the Schechter
parameterization). This can be obtained directly from the redshift
data, (see Loveday et al.\ 1992), but it is better to use the number
counts from the largest available photometric survey. Our photometric
survey covers 10 square degrees, and the number counts have been
confirmed by the results of Huang et al.\ (1997). We therefore
follow Mobasher et al.\ (1993), and determine $\phi^*$ using a
model of the $K-$band number counts based on our estimated values
of $M^*$ and $\alpha$. The model is described in Paper I, but we
have used the distribution of spectral types (and thus star formation
history within the GISSEL96 models) determined by the methods
discussed above. We plot in Figure~\ref{fignc} a compilation of
the $K-$band number counts, along with our model predictions based
upon the values of $M^*$ and $\alpha$ listed in Table~\ref{geomtab}.
The normalization for each model was determined by a least-squares
fit to the number counts from Paper I and is listed in Table~\ref{geomtab}.

\subsection{The Effects of Incompleteness}

We consider here possible biases arising from the 10\% of the
objects for which we attempted spectroscopy, but failed to secure
an identification. We used a two sample Kolmogorov-Smirnov test to
determine whether the identified and unidentified galaxies are
drawn from the same population. The two samples are not different
at the $3\sigma$ level in $I-K$ color or $B-K$ color, but do differ
in their apparent $K$ magnitude distribution since the unidentified
galaxies are mostly at the faint end. They are also different in
$I-$band central surface brightness, but this is due primarily to
the different distributions in apparent magnitude in the two samples.
There was no significant difference between the two samples in the
quantity central surface brightness minus total magnitude, as
measured in the $I$ band. The $K-$band central surface brightness
is more difficult to measure due to the large (2 arcsec) pixels
used, but this is a less relevant quantity since we used optical
spectroscopy to identify the galaxies.

The unidentified galaxies are mainly at the faint end, as every
galaxy with $K<13.25$ was identified. To test whether this apparent
magnitude selection significantly affects the measured luminosity
function, we re-ran the Monte Carlo simulation discussed above,
this time creating a mock catalog of 564 galaxies in each simulation.
From this mock catalog we removed 54 galaxies with the same apparent
magnitude distribution as the unidentified galaxies in our survey.
We repeated this process 1000 times, and the mean results of this
simulation differed from the input parameters by $-0.04 mag$ in
$M^*$ and $+0.04$ in $\alpha$. This is less than the rms statistical
error, but represents the possible systematic error due to
incompleteness. We estimate the total systematic error due to
incompleteness and photometric errors within the spectroscopic
catalog to be less than $0.1 mag$ in $M^*$, and less than $0.1$ in
$\alpha$. The effects of possible incompleteness in the photometric
catalog due to surface brightness or other effects are beyond the
scope of this paper and will be considered elsewhere (Gardner et
al.\ in preparation).

\section{Discussion}

We have presented the first determination of the $K-$band luminosity
function of field galaxies from a wide-field $K-$band selected
spectroscopic redshift survey. Our completeness of 90\% is comparable
to that in optically selected surveys, and we estimate that the
systematic errors due to incompleteness and photometric errors are
smaller than the statistical errors due to the number of galaxies
in our sample.

\begin{figure}

\plotone{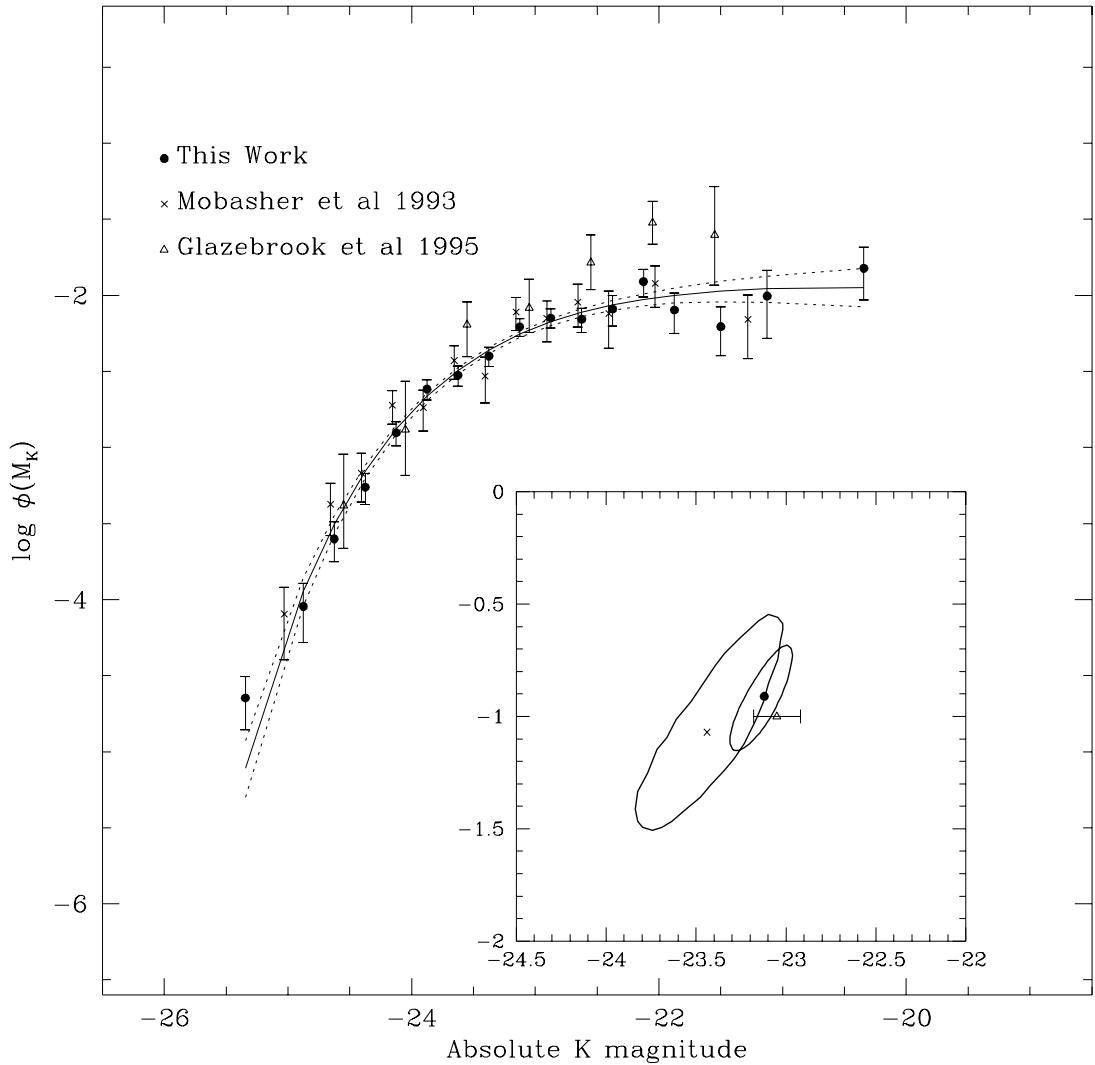}

\caption{Our measurements of the $K-$band luminosity function from
Figure 1 compared to previous measurements by Mobasher et al.\ (1993)
and Glazebrook et al.\ (1995). Inset are the errors in the Schechter
function parameter determinations, showing the best fit values of
Mobasher et al.\ (1993) (cross), our measurement (solid dot), and
the fit of Glazebrook et al.\ (1995) as an error bar since they
fixed the value of $\alpha$ in their determination. We have applied
corrections of $+0.22 mag$ and $-0.30 mag$ to the Mobasher et
al.\ (1993) and Glazebrook et al.\ (1995) measurements, respectively
(see text).}

\label{compare}

\end{figure}

Previous determinations of the $K-$band luminosity function were
based on $K-$band photometry of an optically selected redshift
survey (Mobasher et al.\ 1993) and on a redshift survey of a small
number of galaxies selected in the $K-$band (Glazebrook et al.\ 1995).
In Figure~\ref{compare} we compare our results with these two other
determinations. Following Glazebrook et al.\ (1995), we apply a
correction of $+0.22 mag$ to the Mobasher et al.\ (1993) measurement
to account for their method of calculating K-corrections, and an
aperture correction of $-0.30$ to the Glazebrook et al.\ (1995)
measurement. In the inset to the figure we plot the error ellipse
of the STY determination of the Schechter function parameters for
our measurement of the luminosity function, and the error ellipse
from the measurement of Mobasher et al.\ (1993). Glazebrook et
al.\ (1995) fixed $\alpha$ at $-1.0$ and determined $M^*$; their
error estimate is plotted as an error bar in the inset figure. The
errors of the three determinations overlap at better than the $1
\sigma$ level. All three determinations are consistent with a flat
faint end slope of $-1.0$, similar to that determined from most
optical surveys.

Optical surveys reveal an excess of faint blue galaxies over and
above the number predicted by simple models relating local to
distant observations (e.g. Tyson 1988; Lilly et al.\ 1991). The
flat faint-end slope measured in the local $B-$band luminosity
function of galaxies (EEP; Loveday et al.\ 1992) plays an important
role in this interpretation, for the faint blue galaxies might
otherwise be explained by a local population of intrinsically faint
galaxies. Surveys selected in the $K$ band preferentially study
normal, massive galaxies. The simple passive-evolution number count
models in Figure~\ref{fignc} fit the observed counts well at $K<18$,
and the faint blue galaxy population does not dominate the color
distributions until fainter than this (Gardner 1995). $K-$band
surveys present a different picture from optical surveys.  Instead
of rapid evolution at intermediate or even low redshifts, the counts
and colors of the galaxies making up the $K-$band surveys show only
passive evolution of the old stellar population to $z \sim 0.5$.

The data and software used in this paper are available in electronic
form upon request from the authors.

\section*{Acknowledgments}

We thank the WHT support scientist Brian Boyle, the duty engineers,
Steve Crump and Stuart Barker, and the telescope operator Palmira
Arenaz for their help in keeping the instrument working during the
observing run. We thank Ian Lewis for assisting with the Autofib2
fiber positioner. We thank James Annis, Carlton Baugh, Karl
Glazebrook, Andrew Ratcliffe and Luiz Teodoro for useful discussions.
We acknowledge generous allocations of time on the William Herschel
Telescope, and at the Kitt Peak National Observatory. The WHT is
operated on the island of La Palma by the Royal Greenwich Observatory
at the Spanish Observatorio del Roque de los Muchachos of the
Instituto de Astrof\`isica de Canarias. Data reduction and analysis
facilities were provided by the UK Starlink project. This work was
supported by a PPARC rolling grant for Extragalactic Astronomy and
Cosmology at Durham, with additional funding provided by the Goddard
Space Flight Center. CSF acknowledges a PPARC Senior Research
Fellowship.

\end{document}